\documentclass{ws-procs9x6}

\begin{document}

\title{Singlet Free Energies of a Static Quark-Antiquark Pair}
\author{Konstantin Petrov}

\address{Brookhaven National Laboratory, Upton, NY 11973}

\maketitle

\abstracts{
We study the singlet part of the free energy of a static quark anti-quark ($Q \bar Q$)
pair at finite temperature in three flavor QCD with degenerate quark masses  using $N_{\tau}=4$ and $6$ lattices with Asqtad 
staggered fermion action. We look at thermodynamics of the system around phase transition and study its scaling with lattice spacing and quark masses.
}
\section{Simulations and Results}
Static free energy of a quark-antiquark pair is a frequently used tool to study non-perturbatively the in-medium modification of inter-quark forces. It is calculated as 
the difference in the free energy of the system with static quarks and the same system without them, while the temperature remains constant. Due to colour symmetry 
such quantity contains the singlet and the octet contributions, 
and the usual definition of the free energy is thus referred to as the colour-averaged. This quantity has been extensively in $SU(2)$ and $SU(3)$ gauge theories 
e.g. [\refcite{okacz00}] and full QCD [\refcite{karsch01}].
However, singlet and octet channels were considered in detail only in pure gauge theory [\refcite{attig88,okacz02,digal03}].
In the case of full QCD the first results for singlet and octet free
energy for two flavor QCD have appeared only very recently [\refcite{okacz03a}]. Here we present our results for  3 flavor QCD using the so-called Asqtad staggered fermion action [\refcite{asqtad}]
with two different lattice spacings (corresponding to 
$N_t=4$ and $6$) at three different quark masses. 

Our analysis is to a large extent based
on the gauge configurations generated by the MILC collaboration using
the Asqtad action. Therefore we adopt their strategy for fixing the
parameters which is described in Ref.~\refcite{milc_therm1}. 
We use the most recent value of $r_1$ extrapolated to continuum
and to the physical value of the light quark masses $r_1=0.317$ fm 
to convert the lattice units to temperature.

\begin{figure}
\includegraphics[width=2in]{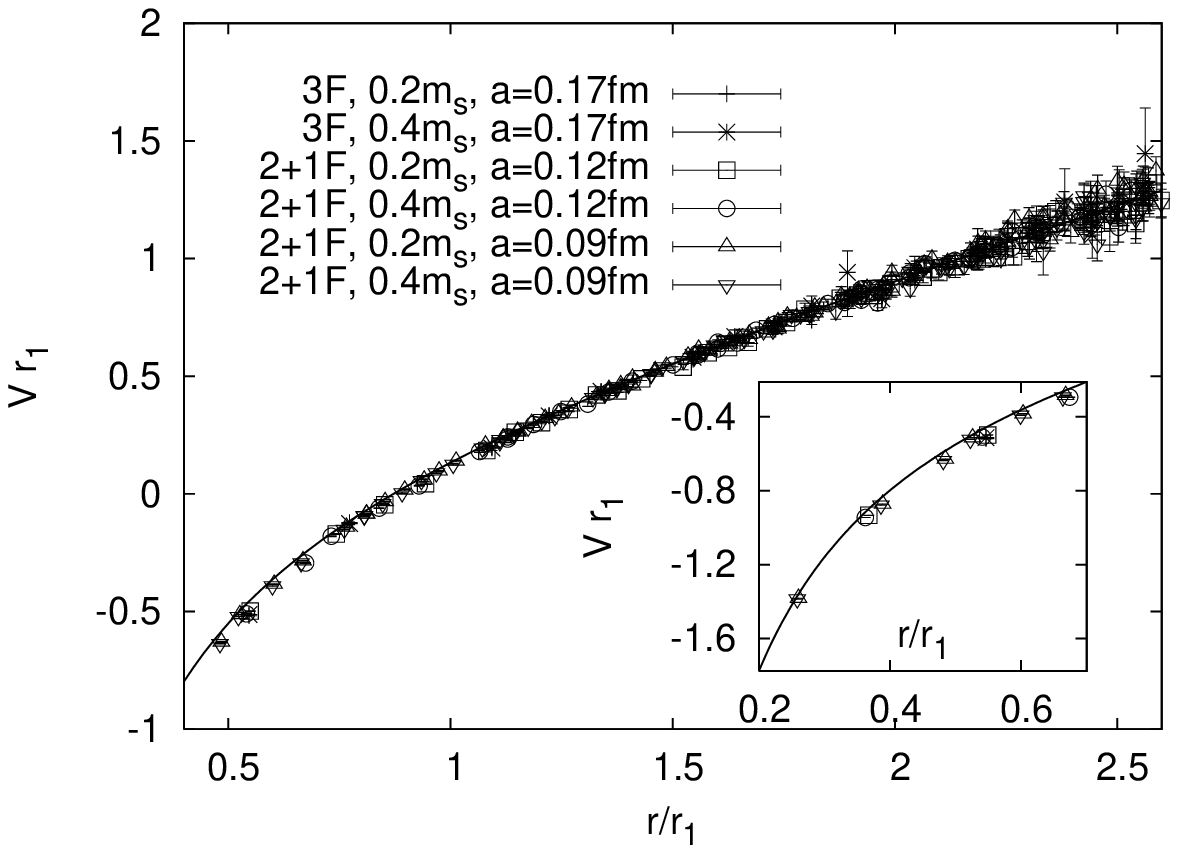}
\includegraphics[width=2in]{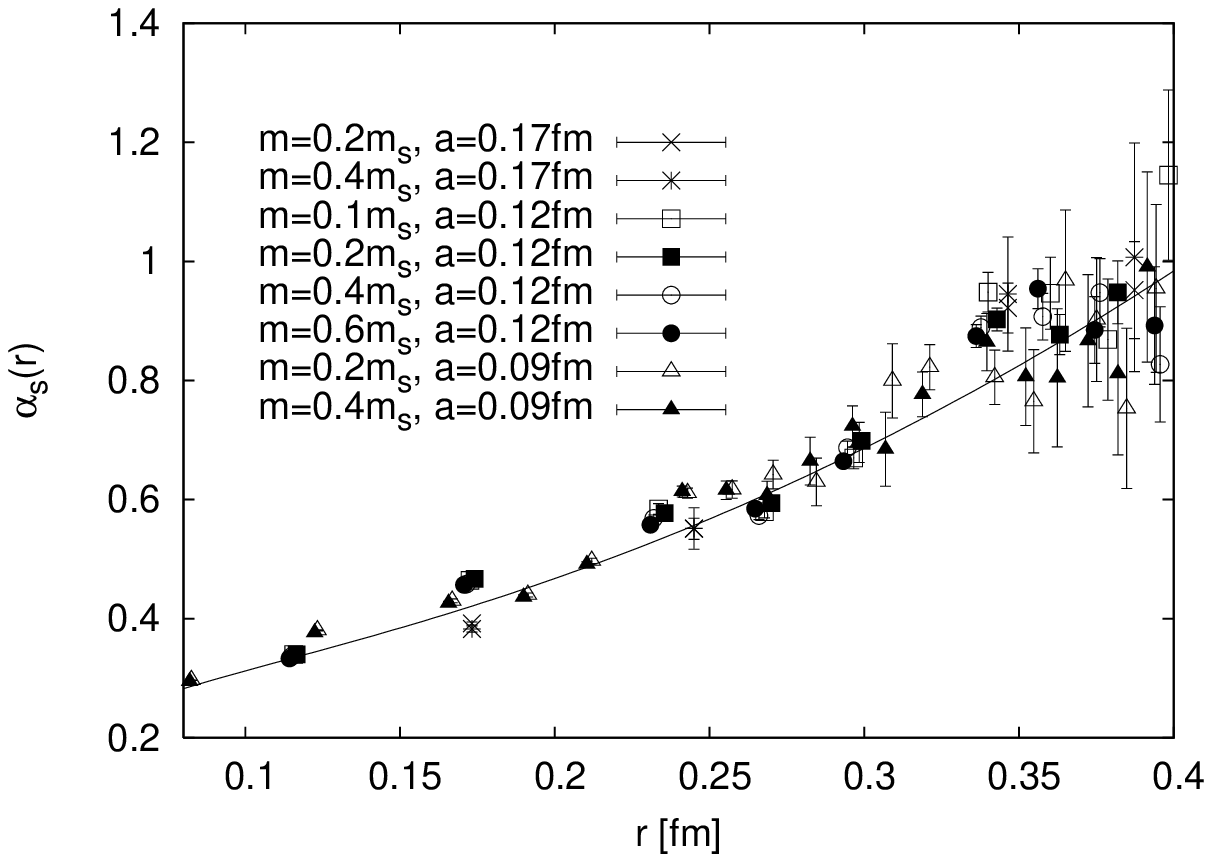}
\label{ztp_fig}
\caption{Zero-temperature static potential (left) and the running coupling (right) [data points from MILC collaboration]}
\end{figure}

The free energy of a static quark anti-quark pair contains a lattice
spacing dependent divergent piece, and thus needs to be renormalized. Following  
Ref.~\refcite{okacz02} we do so by normalizing it to the zero temperature potential at
short distances where the temperature dependence of the free energy 
can be neglected. 
 The static quark potential has been
studied by the MILC collaboration at three different lattice spacings
and various quark masses [\refcite{milc_therm1}].  
We use the following form of the zero temperature
potential, which reproduces MILC data well in the whole range of masses and lattice spacings
\begin{equation}
r_1 V(x)= -\frac{0.44}{x}+0.56 \cdot x+\frac{0.0125}{x^2},~ x=r/r_1
\label{pot_ansatz}
\end{equation}
In Fig. \ref{ztp_fig} we show that both the potential and  effective
coupling constant $\alpha_s(r)$ defined as $\alpha_s(r)=3/4 r^2 dV(r)/dr$  are reproduced well. 
Results presented here come from MILC lattices as well as our simulation for quark masses $m_{light}=0.2m_s, 0.4m_s, 0.6m_s$ on lattices  $12^3 \times 4, 8^3 \times 4$ and $12^3 \times 6$. The temperature range was  $135-412$ MeV for the first two lattice sizes and  $145-310$ MeV for the last one.

Following Ref.~\refcite{mclerran} the free energy of static 
quark-antiquark ( $Q \bar Q$) pair in the color singlet  channels is
defined as
\begin{equation}
\exp(-F_1(r,T)/T+C)=\frac{1}{3} Tr \langle W(\vec{r}) W^{\dagger}(0) \rangle
\end{equation}
Here $W(\vec{x})=\prod_{\tau=0}^{N_{\tau}-1} U_0(\tau,\vec{x})$ is the temporal 
Wilson line, $L(\vec{x})=Tr W(\vec{x})$ is known as the Polyakov loop.
As $W(\vec{x})$ is not gauge invariant one needs to fix a gauge. Here we will use the Coulomb gauge as advocated in [\refcite{ophil02}]. This approach is exactly valid at zero temperature and is numerically true at finite temperature.
One can also consider the color averaged free energy defined as
\begin{equation}
\exp(-F_{av}(r,T)/T+C)= \frac{1}{9} \langle L(\vec{r}) L^{\dagger}(0) \rangle
\end{equation}

\begin{figure}
\includegraphics[width=2in]{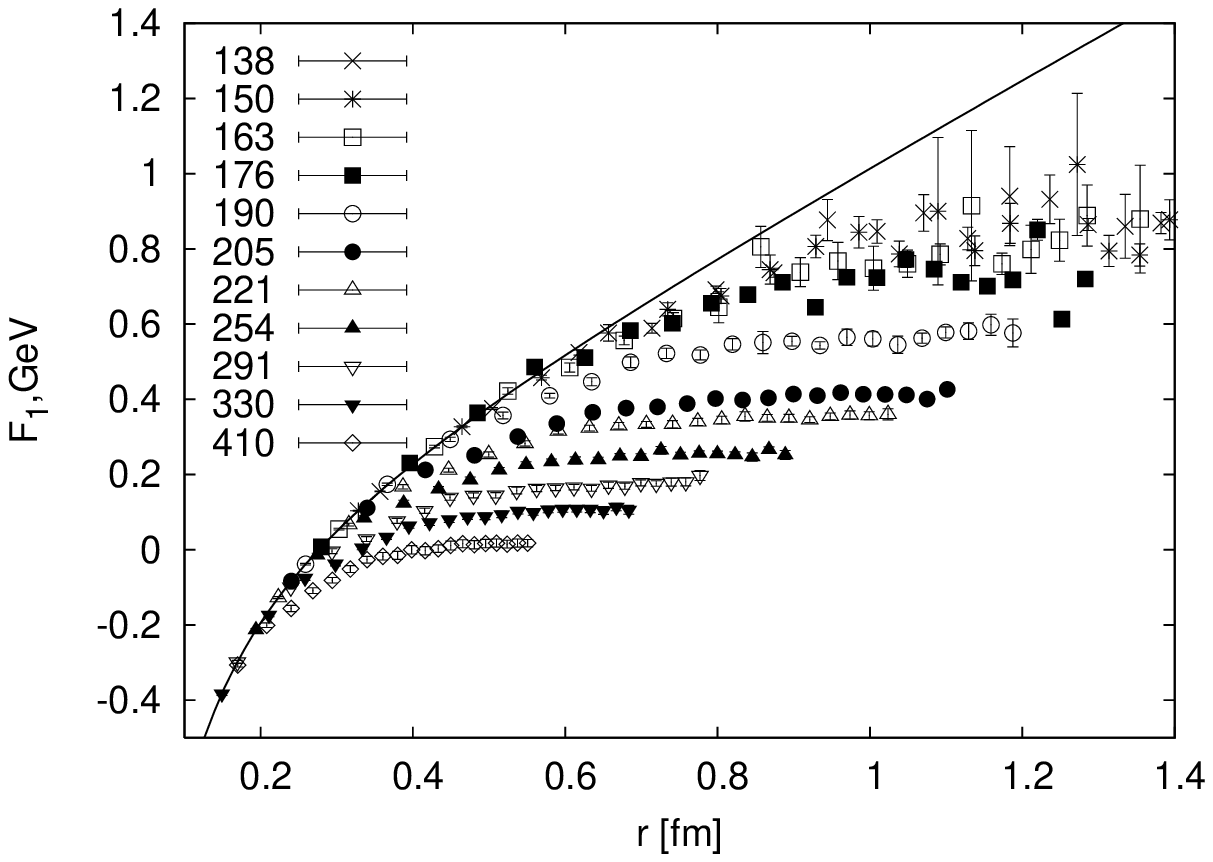}
\includegraphics[width=2in]{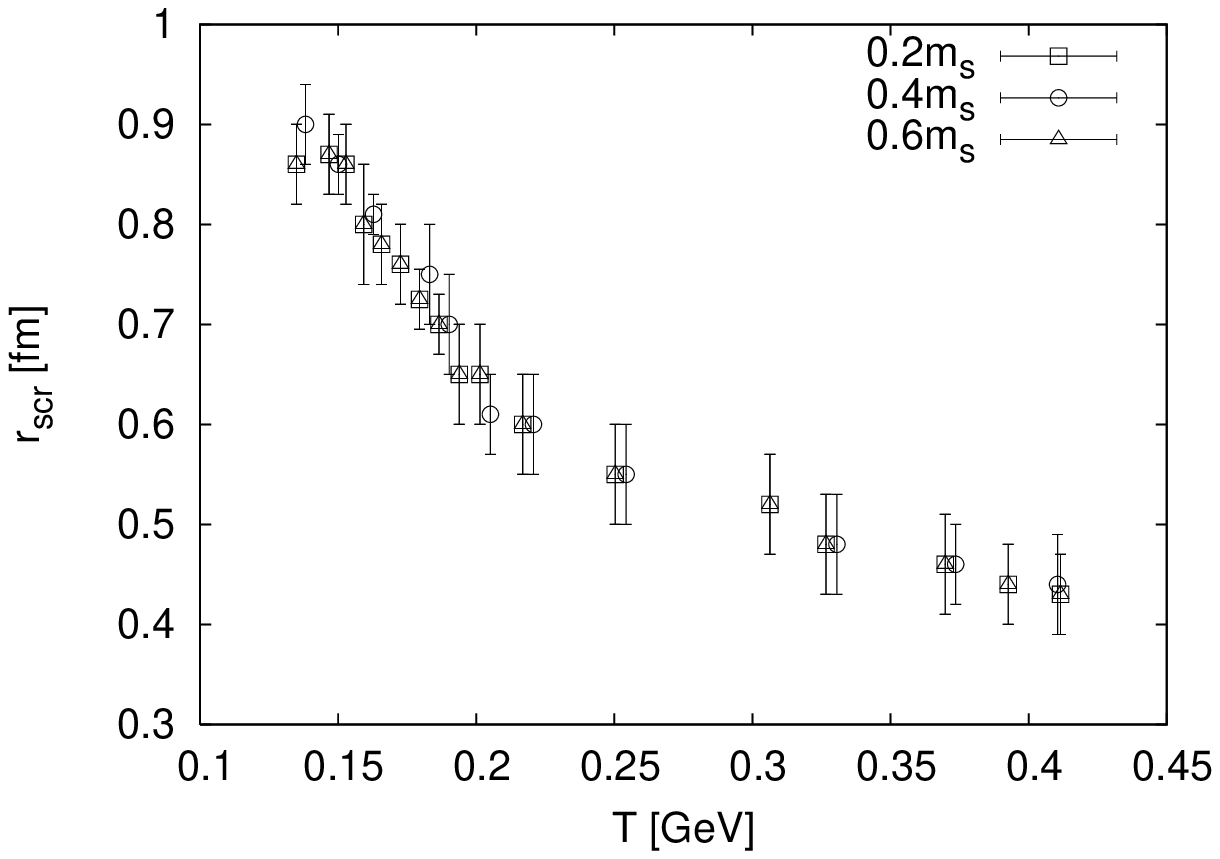}
\caption{Singlet free energies and the effective screening radius }
\label{fig_f04ms_124}
\end{figure}

We start the discussion of our numerical results with the case of the
quark of mass $0.4m_s$ on the $12^3 \times 4$ lattice.
The corresponding numerical results
for the singlet free energy are shown in Fig.\ref{fig_f04ms_124}.
The free energy approaches a finite value $F_{\infty}^i(T)=\lim_{r \rightarrow \infty} F_i(r,T),~~i=1,av$ at large distances which is
usually interpreted as string breaking at low temperature and screening at
high ones. Note that the distance where the free energy effectively flattens
is temperature dependent, it becomes smaller at higher temperatures. At small
distances the singlet free energy is temperature independent and coincides with the
zero temperature potential, as at small distances medium effects
are not important. 
The numerical results for other values of the quark masses are similar. The scaling with the lattice spacing is remarkably good.

To characterize the range of interaction in the medium it is convenient to 
introduce the effective screening radius $r_{scr}$ defined as:
$F_1(r=r_{scr},T)=0.9 F_{\infty}^1(T)$. Here $F_{\infty}^1(T)$ is the asymptotic value
of the singlet free energy at infinite separation. In Fig.\ref{fig_f04ms_124}. we show the
values of $r_{scr}$ for three different quark masses and $12^3\times 4$ lattices.
Certainly as $F_1(r,T)$ has statistical errors it is difficult to determine 
at exactly which distance $r$ the equation $F_1(r=r_{scr},T)=0.9 F_{\infty}^1(T)$ holds.
We have tried to estimate this uncertainty in the values of $r_{scr}$ 
and show them in Fig.\ref{fig_f04ms_124} as errors bars.
At small temperatures the value of the screening radius is about $0.9fm$ and is
temperature independent. As we increase the temperature $r_{scr}$ decreases 
reaching the value of $0.5fm$ at the highest temperature. Note that 
the temperature dependence of $r_{scr}$ is roughly the same for all quark masses.
\begin{figure}
\includegraphics[width=2in]{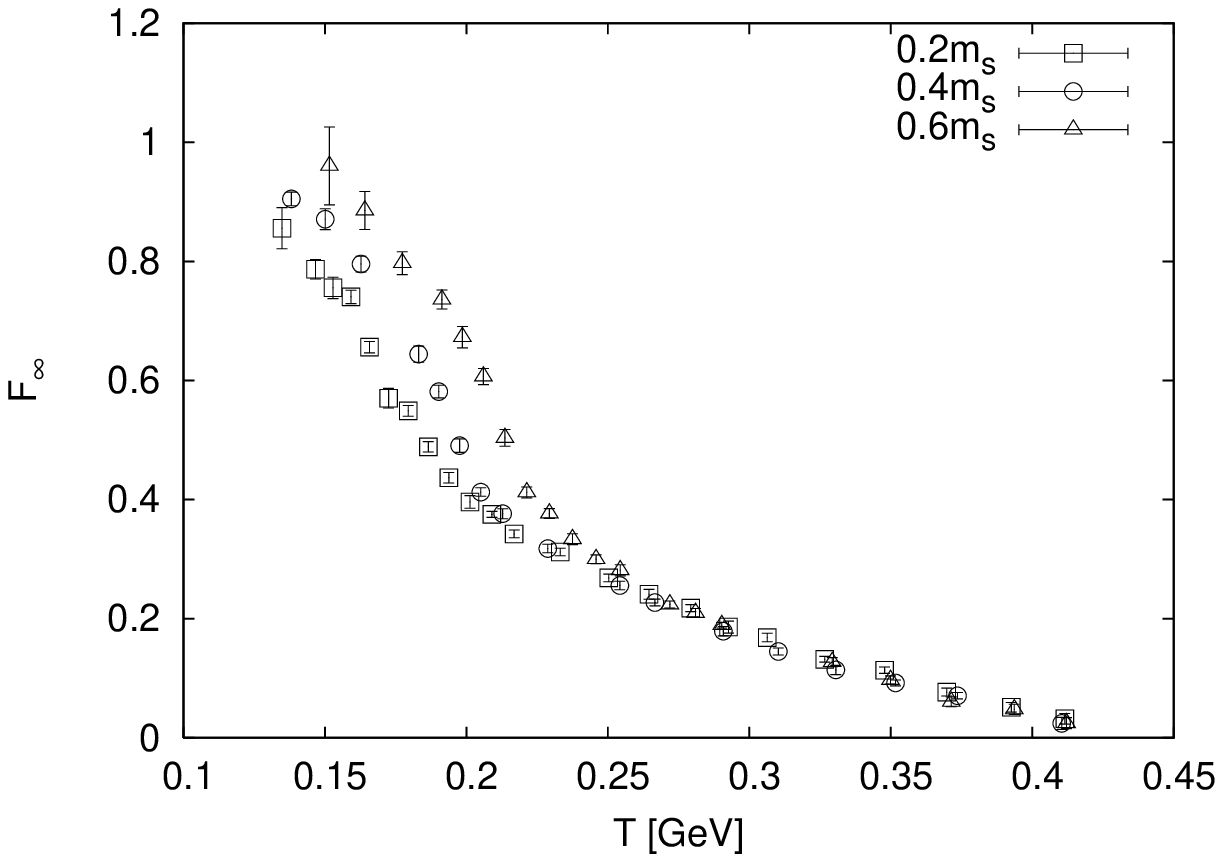}
\includegraphics[width=2in]{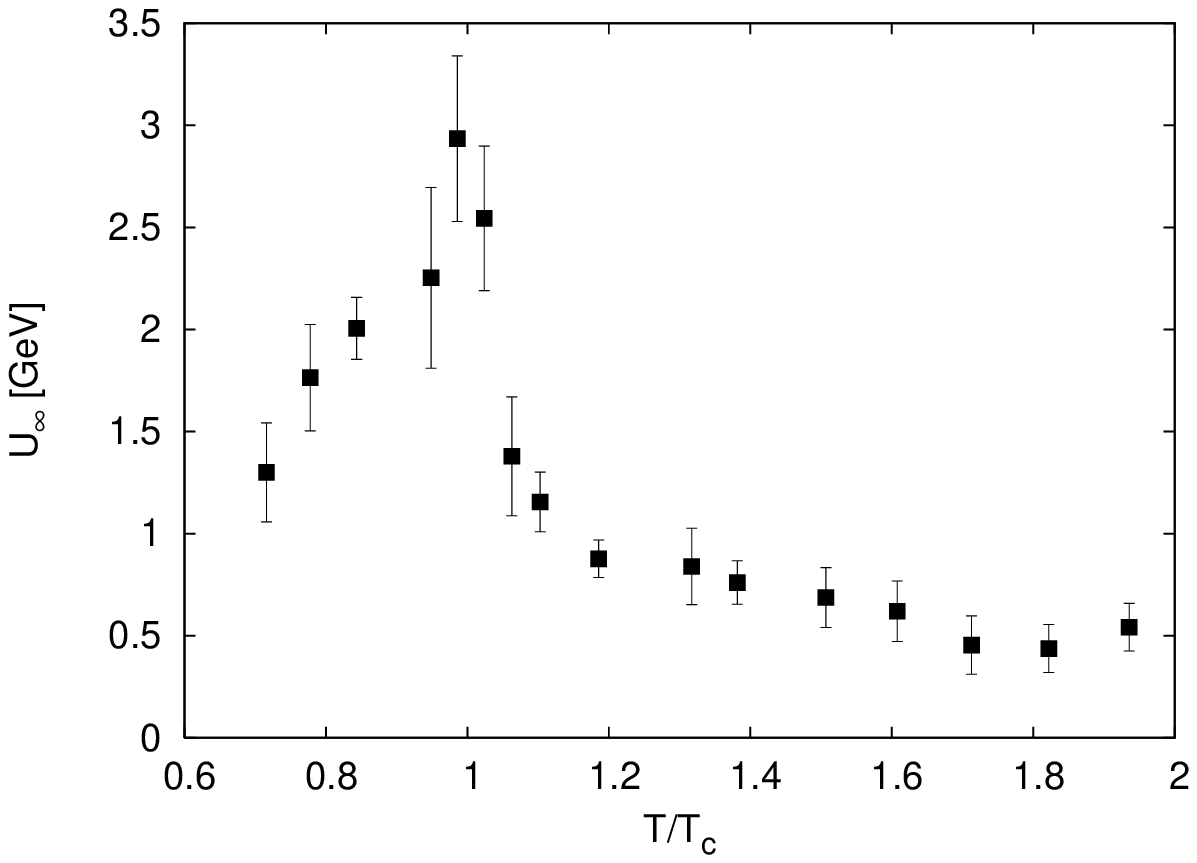}
\label{fig:finf}
\caption{Infinite separation free energies for various quark masses (left) and internal energy at 0.4ms}
\end{figure}
On Fig.3 (left) we plot the asymptotic value of the free energies; the quark 
mass dependence likely vanishes at small temperatures ($T<150MeV$) and definitely
negligible at high temperatures ($T>250MeV$). It is however significant  in the
transition region.

In the limit of very small temperatures we expect 
$F_{\infty}(T)$ to be temperature independent and related to 
twice the binding energy of a heavy-light ($D-$ or $B-$) meson
\begin{equation}
2 E_{bin}=2 M_{D,B}-2m_{c,b}.
\label{ebin_eq}
\end{equation}

Based on this observation in Ref.~\refcite{digal01} it has been argued that
the decrease of $F_{\infty}(T)$ with the temperature close to $T_c$ implies
the decrease of the $M_{D,B}$ leading to quarkonium suppression.
However, $F_{\infty}(T)$ also contains an
entropy contribution due to the presence
of a static $Q \bar Q$ pair:
\begin{equation}
S_{\infty}(T)=-\frac{\partial F_{\infty}(T)}{\partial T}.
\label{sinf_eq}
\end{equation}
Then we can calculate the energy induced by a static quark-anti-quark pair
\begin{equation}
U_{\infty}=F_{\infty}+T S_{\infty}
\label{uinf_eq}
\end{equation}
Numerically the derivative with respect to the temperature in 
Eq. (\ref{sinf_eq}) was estimated using forward differences.

On the right side of Fig.3 we show the energy $U_{\infty}$ as function of temperature. Both
the entropy and the energy show a strong increase near $T_c$.  This large increase in entropy and energy is probably due to many-body effects and makes the
interpretation of $U_{\infty}$ as the binding energy of
heavy-light meson not very plausible.

\section{Conclusions}
The free energy gets screened beyond some distance for all temperatures as 
expected. For small temperature this distance, the effective screening radius,
does not depend on the temperature and is about $0.9fm$. As the temperature
increases the effective screening radius decreases. Light quark mass dependence of
the screening radius is negligible within our statistical accuracy.
 We have also identified the entropy contribution
to the free energy as well as the internal energy at large distances and found that they 
show strong increase at $T_c$. 
We have found substantial quark mass dependence of the free energy in the vicinity of the
transition.

\section*{Acknowledgment}

This work has been supported by the US Department of Energy under
the contract DE-AC02-98CH10886 and by the SciDAC project. It was largely based on the gauge configurations from MILC collaboration; I personally would like to thank Robert Sugar, Doug Toissant, Carleton de Tar and Urs Heller for their help.

\end{document}